\documentclass[preprint,superscriptaddress,floatfix,sort&compress]{revtex4}

\usepackage[dvips]{graphicx}
\usepackage{amssymb,amsfonts,amsmath}
\usepackage{color}
\usepackage[font=scriptsize]{caption}

\renewcommand{\figurename}{Figure}

\usepackage{ulem}

\newcommand{\1}{\begin{equation}}
\newcommand{\2}{\end{equation}}

\if{

\documentclass[twoside,twocolumn,9pt]{article}
\usepackage{extsizes}
\usepackage[super,sort&compress,comma]{natbib} 
\usepackage[version=3]{mhchem}
\usepackage[left=1.5cm, right=1.5cm, top=1.785cm, bottom=2.0cm]{geometry}
\usepackage{balance}
\usepackage{times,mathptmx}
\usepackage{sectsty}
\usepackage{graphicx} 
\usepackage{lastpage}
\usepackage[format=plain,justification=justified,singlelinecheck=false,font={stretch=1.125,small,sf},labelfont=bf,labelsep=space]{caption}
\usepackage{float}
\usepackage{fancyhdr}
\usepackage{fnpos}
\usepackage[english]{babel}
\addto{\captionsenglish}{%
  
}
\usepackage{array}
\usepackage{droidsans}
\usepackage{charter}
\usepackage[T1]{fontenc}
\usepackage[usenames,dvipsnames]{xcolor}
\usepackage{setspace}
\usepackage[compact]{titlesec}
\usepackage{hyperref}

\definecolor{cream}{RGB}{222,217,201}

\begin{document}

\pagestyle{fancy}
\thispagestyle{plain}
\fancypagestyle{plain}{

\fancyhead[C]{\includegraphics[width=18.5cm]{head_foot/header_bar}}
\fancyhead[L]{\hspace{0cm}\vspace{1.5cm}\includegraphics[height=30pt]{head_foot/journal_name}}
\fancyhead[R]{\hspace{0cm}\vspace{1.7cm}\includegraphics[height=55pt]{head_foot/RSC_LOGO_CMYK}}
\renewcommand{\headrulewidth}{0pt}
}

\makeFNbottom
\makeatletter
\renewcommand\LARGE{\@setfontsize\LARGE{15pt}{17}}
\renewcommand\Large{\@setfontsize\Large{12pt}{14}}
\renewcommand\large{\@setfontsize\large{10pt}{12}}
\renewcommand\footnotesize{\@setfontsize\footnotesize{7pt}{10}}
\renewcommand\scriptsize{\@setfontsize\scriptsize{7pt}{7}}
\makeatother

\renewcommand{\thefootnote}{\fnsymbol{footnote}}
\renewcommand\footnoterule{\vspace*{1pt}%
\color{cream}\hrule width 3.5in height 0.4pt \color{black} \vspace*{5pt}} 
\setcounter{secnumdepth}{5}

\makeatletter 
\renewcommand\@biblabel[1]{#1}            
\renewcommand\@makefntext[1]%
{\noindent\makebox[0pt][r]{\@thefnmark\,}#1}
\makeatother 
\renewcommand{\figurename}{\small{Fig.}~}
\sectionfont{\sffamily\Large}
\subsectionfont{\normalsize}
\subsubsectionfont{\bf}
\setstretch{1.125} 
\setlength{\skip\footins}{0.8cm}
\setlength{\footnotesep}{0.25cm}
\setlength{\jot}{10pt}
\titlespacing*{\section}{0pt}{4pt}{4pt}
\titlespacing*{\subsection}{0pt}{15pt}{1pt}

\fancyfoot{}
\fancyfoot[LO,RE]{\vspace{-7.1pt}\includegraphics[height=9pt]{head_foot/LF}}
\fancyfoot[CO]{\vspace{-7.1pt}\hspace{13.2cm}\includegraphics{head_foot/RF}}
\fancyfoot[CE]{\vspace{-7.2pt}\hspace{-14.2cm}\includegraphics{head_foot/RF}}
\fancyfoot[RO]{\footnotesize{\sffamily{1--\pageref{LastPage} ~\textbar  \hspace{2pt}\thepage}}}
\fancyfoot[LE]{\footnotesize{\sffamily{\thepage~\textbar\hspace{3.45cm} 1--\pageref{LastPage}}}}
\fancyhead{}
\renewcommand{\headrulewidth}{0pt} 
\renewcommand{\footrulewidth}{0pt}
\setlength{\arrayrulewidth}{1pt}
\setlength{\columnsep}{6.5mm}
\setlength\bibsep{1pt}

\makeatletter 
\newlength{\figrulesep} 
\setlength{\figrulesep}{0.5\textfloatsep} 

\newcommand{\topfigrule}{\vspace*{-1pt}%
\noindent{\color{cream}\rule[-\figrulesep]{\columnwidth}{1.5pt}} }

\newcommand{\botfigrule}{\vspace*{-2pt}%
\noindent{\color{cream}\rule[\figrulesep]{\columnwidth}{1.5pt}} }

\newcommand{\dblfigrule}{\vspace*{-1pt}%
\noindent{\color{cream}\rule[-\figrulesep]{\textwidth}{1.5pt}} }

\makeatother

\twocolumn[
  \begin{@twocolumnfalse}
\vspace{3cm}
\sffamily
\begin{tabular}{m{4.5cm} p{13.5cm} }

\includegraphics{head_foot/DOI} & \noindent\LARGE{\textbf{Dynamic clustering and re-dispersion in concentrated colloid-active gel composites}} \\
 & \vspace{0.3cm} \\

& \noindent\large{G. Foffano,\textit{$^{a}$} J.~S. Lintuvuori,\textit{$^{b}$} K. Stratford,\textit{$^{c}$} M.~E. Cates,\textit{$^{d}$} and D. Marenduzzo$^{\ast}$\textit{$^{e}$}} \\

\includegraphics{head_foot/dates} & \\

\end{tabular}

 \end{@twocolumnfalse} \vspace{0.6cm}

  ]

\renewcommand*\rmdefault{bch}\normalfont\upshape
\rmfamily
\section*{}
\vspace{-1cm}

\footnotetext{\textit{$^{a}$~Laboratoire de Physique Th\'eorique et Mod\`eles Statistiques, Universit\'e Paris-Sud, UMR 8626, 91405 Orsay, France}}
\footnotetext{\textit{$^{b}$~Univ. Bordeaux, CNRS, LOMA, UMR 5798, F-33405 Talence, France}}
\footnotetext{\textit{$^{c}$~EPCC, School of Physics and Astronomy, Peter Guthrie Tait Road, Edinburgh EH9 3FD, UK }}
\footnotetext{\textit{$^{d}$~DAMTP, Centre for Mathematical Sciences, University of Cambridge, Cambridge CB3 0WA, UK}}
\footnotetext{\textit{$^{e}$~SUPA, School of Physics and Astronomy, University of Edinburgh, James Clerk Maxwell Building, Peter Gutherie Tait Road, Edinburgh EH9 3FD, UK. *E-mail: dmarendu@ph.ed.ac.uk}}

\footnotetext{\dag~Electronic Supplementary Information (ESI) available: [details of any supplementary information available should be included here]. See DOI: 10.1039/b000000x/}

}
\fi

\begin{document}

\title{Dynamic clustering and re-dispersion in concentrated colloid-active gel composites}

\author{G. Foffano}
\affiliation{Laboratoire de Physique Th\'eorique et Mod\`eles Statistiques, Universit\'e Paris-Sud, UMR 8626, 91405 Orsay, France}
\author{J.~S. Lintuvuori}
\affiliation{Univ. Bordeaux, CNRS, LOMA, UMR 5798, F-33405 Talence, France}
\author{K. Stratford}
\affiliation{EPCC, School of Physics and Astronomy, Peter Guthrie Tait Road, Edinburgh EH9 3FD, UK}
\author{M.~E. Cates}
\affiliation{DAMTP, Centre for Mathematical Sciences, University of Cambridge, Cambridge CB3 0WA, UK}
\author{D. Marenduzzo}
\affiliation{SUPA, School of Physics and Astronomy, University of Edinburgh, James Clerk Maxwell Building, Peter Gutherie Tait Road, Edinburgh EH9 3FD, UK. *E-mail: dmarendu@ph.ed.ac.uk}

\begin{abstract}
We study the dynamics of quasi-two-dimensional concentrated suspensions of colloidal particles in active gels by computer simulations. 
Remarkably, we find that activity induces a dynamic clustering of colloids even in the absence of any preferential anchoring of the active nematic director at the particle surface. When such an anchoring is present, active stresses instead compete with elastic forces and re-disperse the aggregates observed in passive colloid-liquid crystal composites. Our quasi-two-dimensional ``inverse'' dispersions of passive particles in active fluids (as opposed to the more common ``direct'' suspensions of active particles in passive fluids) provide a promising route towards the self-assembly of new soft materials.
\end{abstract}

\maketitle



\section{Introduction}
Active fluids are an intriguing example of far-from-equilibrium matter, with various instances present in the living world. Broadly, they consist of suspensions of active particles, which extract energy from their environment and use it to do work~\cite{activefluid,sriram_review,activenematic}. Some well known examples of active fluids include bacterial and algal suspensions, and solutions of cytoskeletal gels interacting with molecular motors, such as actomyosin, or active microtubule networks~\cite{sanchez,dogic}.

The non-thermal forces that active particles exert on their environment can be modelled, in the simplest approximation,  as force dipoles~\cite{activefluid,Ramaswamy_PRL_2004}. 
The interplay between the dynamics of the nematic order parameter describing the orientation of such dipoles and the Navier-Stokes equation, which models the flow of the underlying solvent in the presence of the active forcing, leads to interesting physics. A well-known example is the ``spontaneous flow'' instability, which sets in when the density of active forces becomes large enough~\cite{spontaneousflow1,spontaneousflow2,spontaneousflow3}, and eventually leads to chaotic patterns~\cite{julia,cristina,activeturbulence1,activeturbulence2} not unlike those observed in the ``bacterial turbulence'' exhibited by dense films of {\it B. subtilis}~\cite{bacterialturbulence}.
Other striking examples of activity-induced phenomena in these materials are negative drag in micro-rheology~\cite{negativedrag}, and spontaneous shear banding~\cite{rheoactive1,rheoactive2}.

Here we study what happens in a {{quasi-two-dimensional}} concentrated dispersion of passive colloidal particles within an active fluid. Most work on colloidal suspensions in active systems has focused on the very dilute regime~\cite{tracers1,tracers2,tracers3}. 
Our emphasis in this work is, instead, on the concentrated regime, where fluid-mediated interparticle interactions dominate, and give rise to many-body effects. 
In the related case where colloids are dispersed at high volume fraction within {\it passive} liquid crystals, such many body effects provide an effective avenue to create new soft composite materials, such as colloidal crystals~\cite{bpcolloids,kevinchol,miha,smalyukh}, and self-quenched glasses~\cite{tiffany}.
We find that, when there is no (or weak) anchoring of the active force dipole orientation at the particle surface, activity promotes transient dynamic inhomogeneities in passive particle-active gel mixtures. Remarkably, in the case of strong (normal or tangential) anchoring, activity has the opposite effect; it leads to the partial re-distribution of colloidal aggregates observed in passive colloid-nematic mixtures (which are kept together by elastic forces). Superficially, the conclusion that activity may have competing effects is similar to that reached in~\cite{chantal,brader} for ``direct'' active suspensions (active particles in a passive fluid). However, here we consider an ``inverse'' system (passive particles in an active fluid), and the physical mechanisms underlying our observations are completely distinct. 
Overall, the morphology and physical behaviour of our colloid-active gel composites can be controlled by tuning either anchoring, initial conditions, or the nature of the active host ({\it i.e} extensile or contractile): these mixtures therefore provide a promising route to creating new self-assembled active materials.

\section{Simulation method}

The hydrodynamics of an active nematic fluid can be described by a set of continuum equations~\cite{sriram_review,Ramaswamy_PRL_2004} that govern the time evolution of the velocity field $u_\alpha$ and of a (traceless, symmetric) tensor order parameter $Q_{\alpha\beta}$. The latter describes the orientational order of the active particles (whether bacteria, algae, or cytoskeletal filaments) which usually have a rod-like shape and are thus capable of nematic alignment~\cite{sriram_review}. Without activity, nematics are described by a Landau -- de Gennes free-energy ${\cal F}$, whose density equals $f=F(Q_{\alpha\beta}) + K(\partial_{\beta}Q_{\alpha \beta})^2/2$, with \begin{equation}\label{eq:fed_bulk}
F(Q_{\alpha\beta}) = \frac{A_0}{2}\left(1-\frac{\gamma}{3}\right)Q_{\alpha \beta}^2-\frac{A_0\gamma}{3}Q_{\alpha \beta}Q_{\beta \gamma}Q_{\gamma \alpha} + \frac{A_0\gamma}{4}(Q_{\alpha \beta}^2)^2
\end{equation}
where indices denote Cartesian coordinates, summation over repeated indices is implied, $A_0$ is a free energy scale, $\gamma$ controls the magnitude of nematic order, and $K$ is an elastic constant.

The hydrodynamic equation for the evolution of the order parameter is:
$D_t Q_{\alpha \beta} = \Gamma H_{\alpha \beta}$,
with $D_t$ a material derivative describing advection by the fluid  velocity $u_\alpha$, and rotation/stretch by flow gradients (see \cite{spontaneousflow2}). The molecular field is $H_{\alpha \beta}= -{\delta 
{\cal F} / \delta Q_{\alpha \beta}} + (\delta_{\alpha \beta}/3) {\mbox {\rm Tr}}({\delta {\cal F} / \delta Q_{\alpha \beta}})$ 
and $\Gamma$ is a collective rotational diffusivity. 
The fluid velocity obeys $\partial_\alpha u_\alpha = 0$, and also the Navier-Stokes equation, in which a passive thermodynamic stress enters~\cite{spontaneousflow2}. An additional stress term captures the action of active force dipoles:
\begin{equation}\label{activestress}
\Pi_{\alpha\beta}= 
-\zeta Q_{\alpha \beta}
\end{equation}
where $\zeta$ is the activity parameter that sets the dipolar force density~\cite{Ramaswamy_PRL_2004} ($\zeta<0$ for contractile fluids and $\zeta>0$ for extensile ones).
We solve the Navier-Stokes equation via lattice Boltzmann (LB), and the equation for the order parameter via finite difference~\cite{Juho} methods. 
All the results presented here were obtained in a ``quasi-two-dimensional'' box ($128\times 128\times 16$) with periodic boundary conditions, considering a colloidal volume fraction $\varphi=15$\% (corresponding to $771$ colloids in the simulation box).

{{We aim to study a quasi-2D geometry for a twofold reason. First, the results and patterns are much easier to visualise and interpret in this geometry. Second, this setup allows us to refer to existing simulations of active nematics~\cite{spontaneousflow2,spontaneousflow3,julia,activeturbulence1,activeturbulence2}, which generally studied 2D or quasi-2D systems, where there is translational invariance along one of the directions. Here, we therefore wish to add isometric particles to the simulation in a way which retains the quasi-2D geometry. There are then at least two ways to achieve this. One is to have disks in a plane of 3D nematics, which could be realised by doing LB with cylinders on a finite slab thickness where both the cylinders and the LB obey periodic boundary conditions in the thin direction\footnote{In our cases, boundary conditions are periodic also along the other two directions, but this need not be true in general for a quasi-2D geometry.}. However the types of defect structure induced in a 3D nematic by a cylinder and a sphere are not the same. A second choice, which we adopt here and which avoids this issue, is to drop the strict translational invariance in the normal direction and allow real spheres to be present, but still in a thin sample, where we retain periodic boundary conditions along the normal dimension. If the latter is only a few times the particle radius (to avoid obvious artifacts), then this choice leads to a quasi-2D geometry, with approximate rather than strict translational invariance along the thin direction: for a similar reason, the thinner films studied in~\cite{shendruk2018} are quasi-2D. While fully 3D simulations are outside the scope of the current work, selected simulations in a cubic box lead to qualitatively similar results as found with the quasi-2D geometry studied here.}}

{{Simulations were run with $A_0=1$, $K=0.05$, $\Gamma=0.3$, $\xi=0.7$ (corresponding to a flow-aligning liquid crystal in the passive limit), and $\gamma=3.1$. The choice of $\gamma$ ensures that the absolute minimum of the free energy corresponds to the nematic phase.}}

{{Within our LB scheme, we introduce spherical colloidal particles as solid objects with a velocity no-slip boundary condition at their surface through the standard method of bounce-back on links~\cite{Juho,ladd,Juho2}. 
To impose normal or planar anchoring of the nematic director field -- a headless unit vector $n_\alpha$ oriented along the major principal axis of $Q_{\alpha\beta}$ -- we include a surface free energy density, $f_{\rm s}=\frac{W}{2}(Q_{\alpha\beta}-Q_{\alpha\beta}^0)^2$, favouring a suitable value of the tensorial order parameter, $Q_{\alpha\beta}^0$, at the colloidal surface~\cite{Juho,Juho2}. The quantity $W$ is the strength of the anchoring. The total free energy density leads to the following boundary conditions which we impose on the surface of the colloidal spheres~\cite{Juho2},
\begin{equation}\label{bc}
\nu_{\gamma}\frac{\partial f}{\partial \partial_{\gamma}Q_{\alpha\beta}}+\frac{\partial f_{s}}{\partial Q_{\alpha\beta}}=0,
\end{equation}
where $\nu_{\gamma}$ is the local outer normal to the colloidal surface. Eq.~\ref{bc} is the way in which anchoring enters our algorithm in practice.}

{{We find that the impact of colloids depends strongly on the dimensionless ratio $w=WR/K$ where $W$ is the anchoring strength of the nematic director at the particle surface, and $R$ is the particle radius. 
The value of $w$ determines how much colloidal particles affect the orientational order of the liquid crystal nearby. In our simulations we use $w=0$, corresponding to no preferred anchoring, or $w=2.3$, which corresponds to strong anchoring -- either hometropic/normal, or degenerate planar. }}

{In our simulations the colloidal particles have a hard sphere radius equal to $R=2.3$ lattice sites (the only exception is Fig.~4 where $R=5.3$). To prevent particle overlap, we include an additional short range soft potential, of the form $V(h)=\epsilon \exp\left[-(h/\sigma)^{\nu}\right]$, where $h$ is the surface-to-surface separation between any two particles, and other parameters are set as $\epsilon=0.004$, $\sigma=0.8$, $\nu=6$. The potential and its resulting force are both set to zero at a cutoff distance $h_c=1.5$. 
The values of $\nu$ and $h_c$ are larger than in previous work~\cite{kevinchol} as with normal anchoring particles attract quite strongly due to the presence of defects, and this may lead to overlaps with a softer potential. 
To quantify clustering in Figures~2 and 5, we considered that two colloidal particles belong to the same cluster if their centre-to-centre distance is smaller than a binding distance $d_b$, which we took slightly larger than $2R+h_c$ (we checked that the trends we see do not depend on the exact value of $d_b$). For more details on our simulation method see~\cite{kevinchol,Juho}.}

\begin{figure}[!h]
  \begin{center}
    \includegraphics[width=1.1\columnwidth]{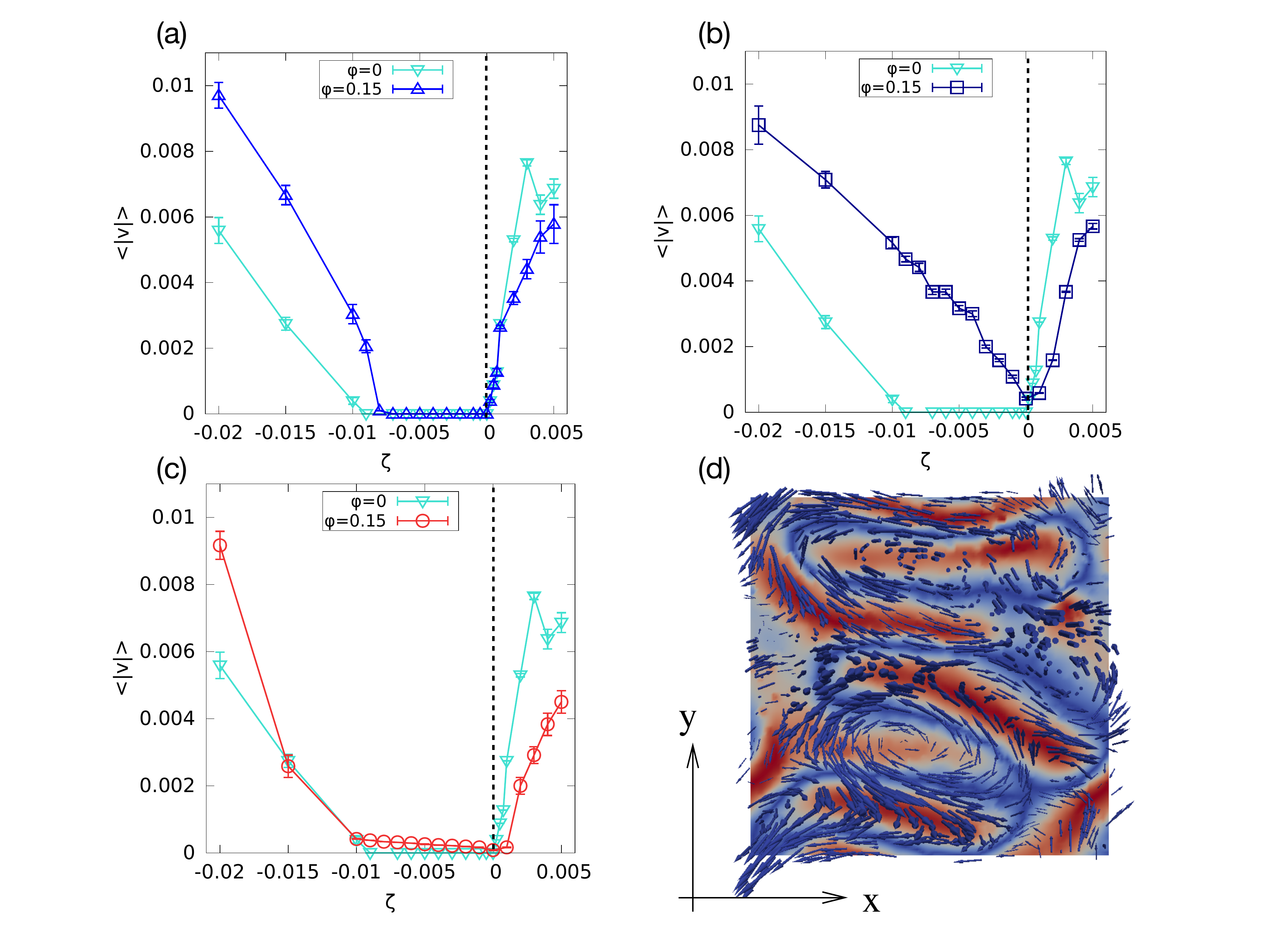}
    \caption{The average modulus of the fluid velocity is studied as a function of the activity $\zeta$ for systems with colloidal volume fraction $\varphi = 15$\% in the cases of: (a) no anchoring ($w=0$), (b) normal anchoring ($w=2.3$) and (c) planar anchoring ($w=2.3$) of the liquid crystal director at the colloidal surfaces. All the results are compared to the case without particles (right-pointing triangles in (a,b,c)). Panel (d) shows an example of director orientation and fluid velocities in the spontanously flowing state with $\zeta=0.005$ (extensile) and $w=0$. Here the colour coding refers to the orientation of the director with blue along $x$ and red perpendicular. Arrows denote fluid flow.}
\end{center}
\end{figure}


\section{Results}

We first study the effects of a { quasi-2D} concentrated colloidal dispersion on the onset of spontaneous flow within the active gel, which is triggered by an increase in the activity $\zeta$ -- this problem has been well studied in the absence of colloids~\cite{spontaneousflow1,spontaneousflow2,spontaneousflow3}. 
In Fig.1(a-c) we measure the fluid average velocity  $\langle |v|\rangle$ as a function of $\zeta$, and compare this to the curve recorded when no colloids are present (Fig. 1; right-pointing triangles). Note that, in these simulations (as well as in Fig. 2 below), the liquid crystal host is initialised in the ordered phase (along $x$ in Fig. 1). 

In the weak anchoring regime ($w\rightarrow 0$) colloidal particles do not perturb the nematic order in the static limit; the spontaneous flow patterns resemble those found with $\varphi=0$~\cite{spontaneousflow1,spontaneousflow2,spontaneousflow3}, and at large $|\zeta|$ they feature vortices associated with splay-bend distortions in the orientational order (Fig. 1(d), Suppl. Movie 1). 
Colloids significantly affect, instead, the {\it magnitude} of the flow (Fig. 1(a)). Interestingly the behaviour is opposite for contractile and extensile activities: in contractile nematics, colloids cause an {\it increase} of $\langle  |v|\rangle$; instead, in extensile suspensions particles lead to a {\it decrease} in the magnitude of spontaneous flow (Fig. 1(a)). 
This different behaviour is possibly related to an excess of splay with respect to bend deformation created by the colloids as they move through the active host~\cite{noanchsplay}: contractile nematic fluids are especially sensitive to splay~\cite{sriram_review} (extensile respond more strongly to bend), and this could
 explain our observation. 

For finite anchoring ($w= 2.3$), the colloidal inclusions have a much stronger effect on the stability of a quiescent fluid, also at small $|\zeta|$. 
For particles in a contractile gel with normal anchoring, there is now barely a passive, quiescent phase -- a global spontaneous flow appears for very small $|\zeta|$ (Fig. 1(b)). In stark contrast, if the anchoring is planar only a weak flow localised around the particles sets up for small activities (Fig. 1(c)).
 The very different behaviour of the two contractile suspensions is linked to the distinct defect patterns created around the colloidal particles: normal anchoring leads to the formation of extended disclinations hugging the colloidal surface, whereas planar anchoring favours the appearance of localised point defects~\cite{lcreview}.
As elastic distortions, which are also associated with defects, drive active stresses and flows (see {\it e.g.} Eq.~\ref{activestress} and Ref.~\cite{sriram_review}), the latter are naturally stronger in the normal anchoring case, where disclinations are larger in size.

\begin{figure*}
  \begin{center}
    \includegraphics[width=1.0\textwidth]{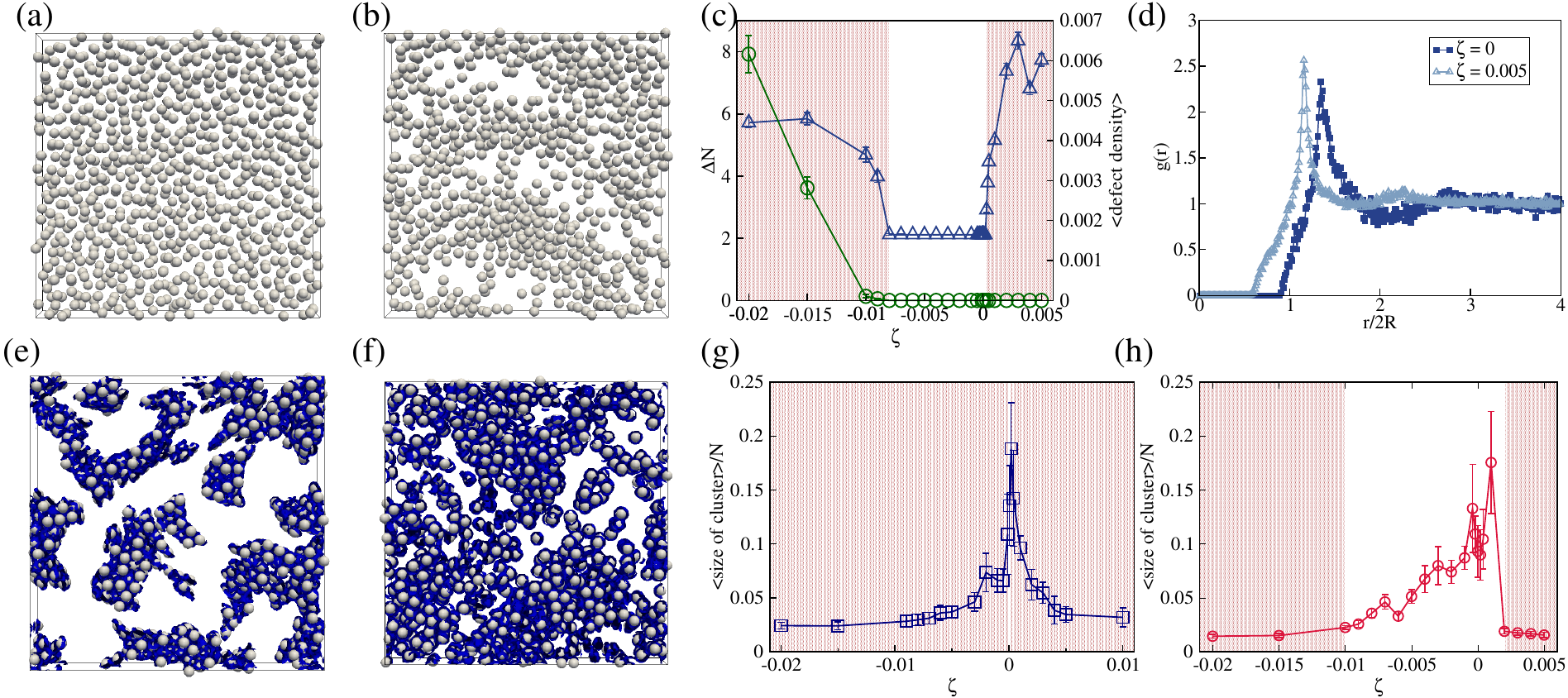}
  \end{center}
  \caption{Simulation results for a system consisting of $\varphi = 15$\% volume fraction of colloidal particles, with free (first row, $w=0$) and normal (second row, $w=2.3$) anchoring. First and second columns show snapshots of particles in passive ($\zeta=0$) and active (extensile) nematic fluids ($\zeta=0.005$ for (b) and $\zeta=0.01$ for (f)). The dark regions in (e) and (f) denote the presence of defects due to normal anchoring. 
 Panel (c) shows that a sudden increase in $\Delta N$ (triangles) at $w=0$ is correlated with the onset of spontaneous flow (shaded regions refer to values of $\zeta$ where $\langle|v|\rangle>10^{-4}$: full curves are presented in Fig.1(a)) and not with an increase in the extent of disclinations (circles are used to show the values of the defect density, computed as the volume fraction of regions where the scalar order parameter of the liquid crystal is below 0.45 -- 0.5 refers to a perfectly aligned nematic). $|\Delta N|$ was computed by binning the simulation into $16\times 16\times 16$ cubes. Panel (d) shows the pair correlation function $g(r)$ for various $\zeta$ at $w=0$. Panels (g) and (h) show the averaged cluster size as a function of $\zeta$ for normal and planar anchoring respectively {{(this is normalised by the total number of particles $N$)}}. Shaded regions are as in panel (c).}
\end{figure*}

Having characterised the dynamics of the active fluid, we now examine the colloids. One of our main results is that, even at $w=0$, activity in the nematic host drives clustering of particles (Fig. 2(a)-(d), Suppl. Movie 1). Such clusters are not long-lived: they are dynamic and keep exchanging particles, so that higher order peaks are absent in the pair correlation function, $g(r)$ (Fig. 2(d); {{note this is calculated from the $2D$ projection of colloidal centres in keeping with our quasi-2D geometry philosophy}}). Instead, the dynamic clustering leaves a detectable signature in the fluctuations of the particle number density, $\Delta N=\sqrt{\langle N^2\rangle - \langle N\rangle ^2}$ {{in a $16\times 16 \times 16$ cube, Fig. 2(c))~\cite{giantdensityfluctuations}}}. These fluctuations correlate well with the overall inhomogeneous colloidal distribution, and show a clear increase at the onset of spontaneous flow in the host (Fig. 2 (c), shaded region), while there is no direct correlation with the presence of defects (Fig. 2(c)). 
This observation suggests that the mechanism leading to the observed clustering is not through defect-mediated or elastic interactions, but rather through advective forces from the spontaneous flow in the active gel. These forces endow the particles with a nonequilibrium and spatially inhomogeneous effective mobility which may promote their aggregation. It is important to stress that the viability of this as a possible route to the dynamic clustering, relies crucially on the non-equilibrium nature of spontaneous flow: a Brownian diffusivity, even if space dependent, can never lead to an inhomogeneous particle distribution. In this respect the dynamic clustering may be viewed as a distant relative of motility-induced phase separation in suspensions of self-propelled particles~\cite{mips,buttinoni,palacci1}, although both the underlying physics and the emerging dynamics are very different. {{More similar physics to the case at hand is that underlying fluctuation-dominated phase ordering~\cite{barma,barma2}, or equivalently path coalescence, which leads to aggregation of pointlike particles in random, turbulent, flow~\cite{mehlig,amos}. As in those systems, in our anchoring-free suspensions clustering is mainly due to the fact that particles are subjected to the same history (here, the same active turbulent flows, which can be viewed as a noisy forcing term with spatiotemporal correlations). In our colloid-active gel composites, excluded volume and near-field effects play additional roles.}}

When $w>0$, the situation is very different. Due to the finite anchoring, colloids disrupt the nematic ordering nearby and create defects close to their surface, which can interact at high volume fractions. If two or more colloids come together, they share their disclinations, and this reduces the free energy cost of elastic distortions in the nematic fluid. This drives thermodynamic and stable particle clustering at $\zeta=0$ (Fig. 2 (e)). 
Remarkably, activity now disrupts this clustering: spontaneous flow breaks the aggregates and re-disperses particles, leading to a more uniform distribution in steady state (Fig. 2(f)). This can be seen from the reduction of the average cluster size (Fig. 2(g)) as well as from the smearing out of the higher order neighbour peaks in $g(r)$ (Fig.~\ref{fig:1-SI}). The situation with planar anchoring is similar (Fig. 2(h) and Fig.~\ref{fig:1-SI}).

Most of the results presented up to now all hinge on the existence of a spontaneous flow in the active nematic. To gain a more complete understanding of the effects of activity on the colloid-active gel mixture, it is also of interest to investigate the effective interparticle interactions in a defect-free active nematic background, and in the absence of global spontaneous flow.
In Figure~\ref{fig:2-SI}, we present the results of further simulations where we investigate the character of the effective interaction between two colloids embedded in active nematics. Here the particles were initially at a distance large enough to prevent any attraction at $\zeta=0$, and we chose large values of $\zeta$ so as to enhance the effect of the local flow that generates due to disclinations at the colloidal surface, before a chaotic flow could set up in the bulk. For each type of anchoring (normal or planar) we considered the cases where the dimer is oriented either parallel or orthogonal to the far field director (here directed along $x$). Therefore there are four possible cases for contractile gels, and another four for extensile ones.  
We found that, for both contractile and extensile gels, in three cases out of these four the colloids were moving away from each another (see red arrows in Fig.~\ref{fig:2-SI}), denoting a general tendency to repulsion.
The case in which this study is most relevant in the one of planar anchoring, when $-0.01<\zeta<0$, as apparent from Fig.~1(d) of the main text. There colloids tend to orient at about 30 degrees (see e.g.~\cite{lcreview}),
at $\zeta=0$, and local flow generated a competition between attraction and repulsion (g,h), responsible for the slow decrease in the average cluster size evident in Fig.~2(h) for that range of $\zeta$.

\begin{figure}[ht]
\includegraphics[width=\linewidth]{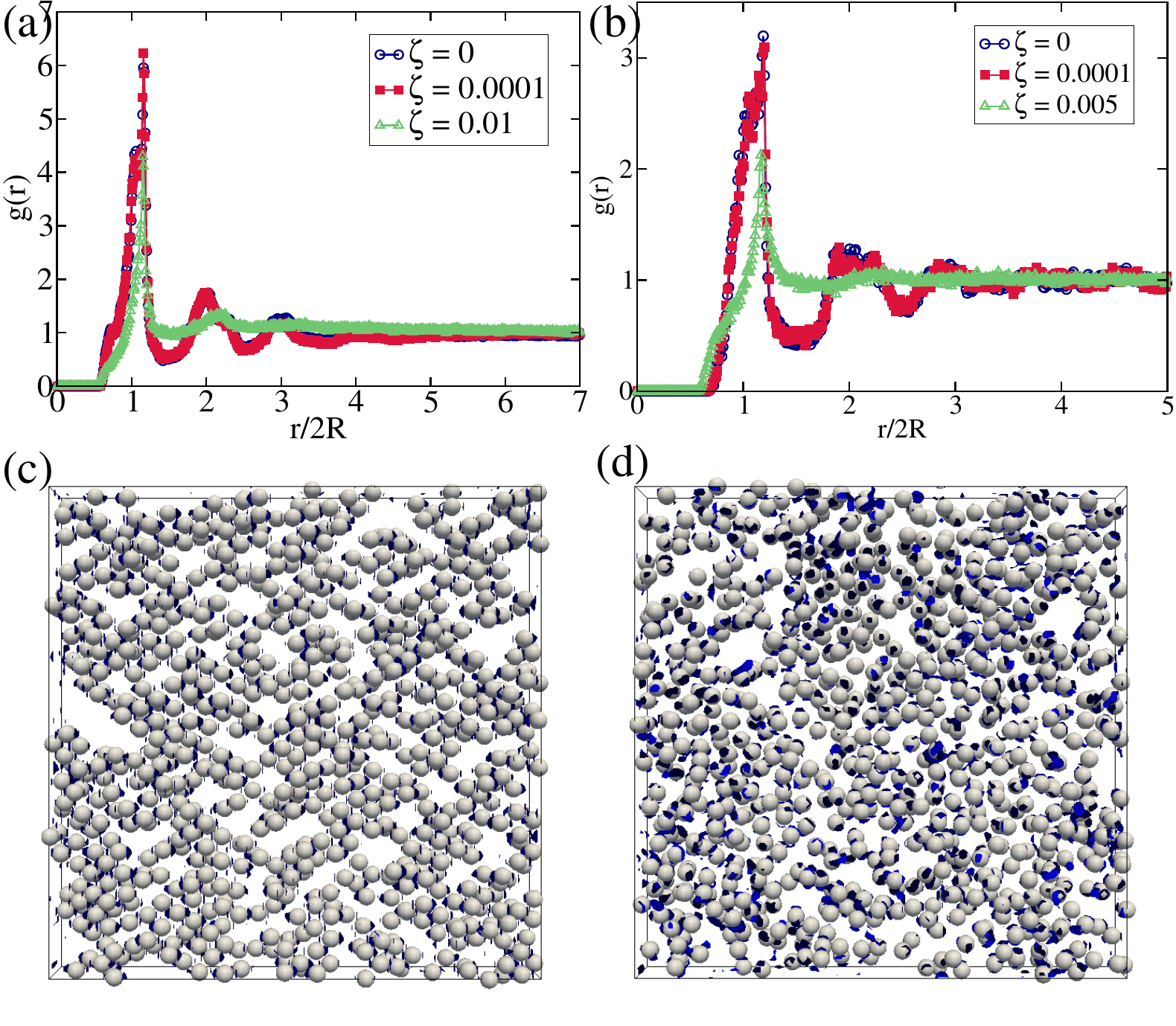}
\caption{In the first row we show the pair correlation function $g(r)$ for a system consisting $\varphi = 15$\% volume fraction of colloidal particles, in extensile active liquid crystals, for various $\zeta$ (see legend), when (a) normal or (b) planar anchoring is imposed at the colloidal surface. Panels (c) and (d) are snapshots of particles in passive ($\zeta=0$ (a)) and active (extensile) host phases ($\zeta=0.01$ (b)), in the case of {\it planar} anchoring.}
\label{fig:1-SI}
\end{figure}

\begin{figure}
\includegraphics[width=0.8\linewidth]{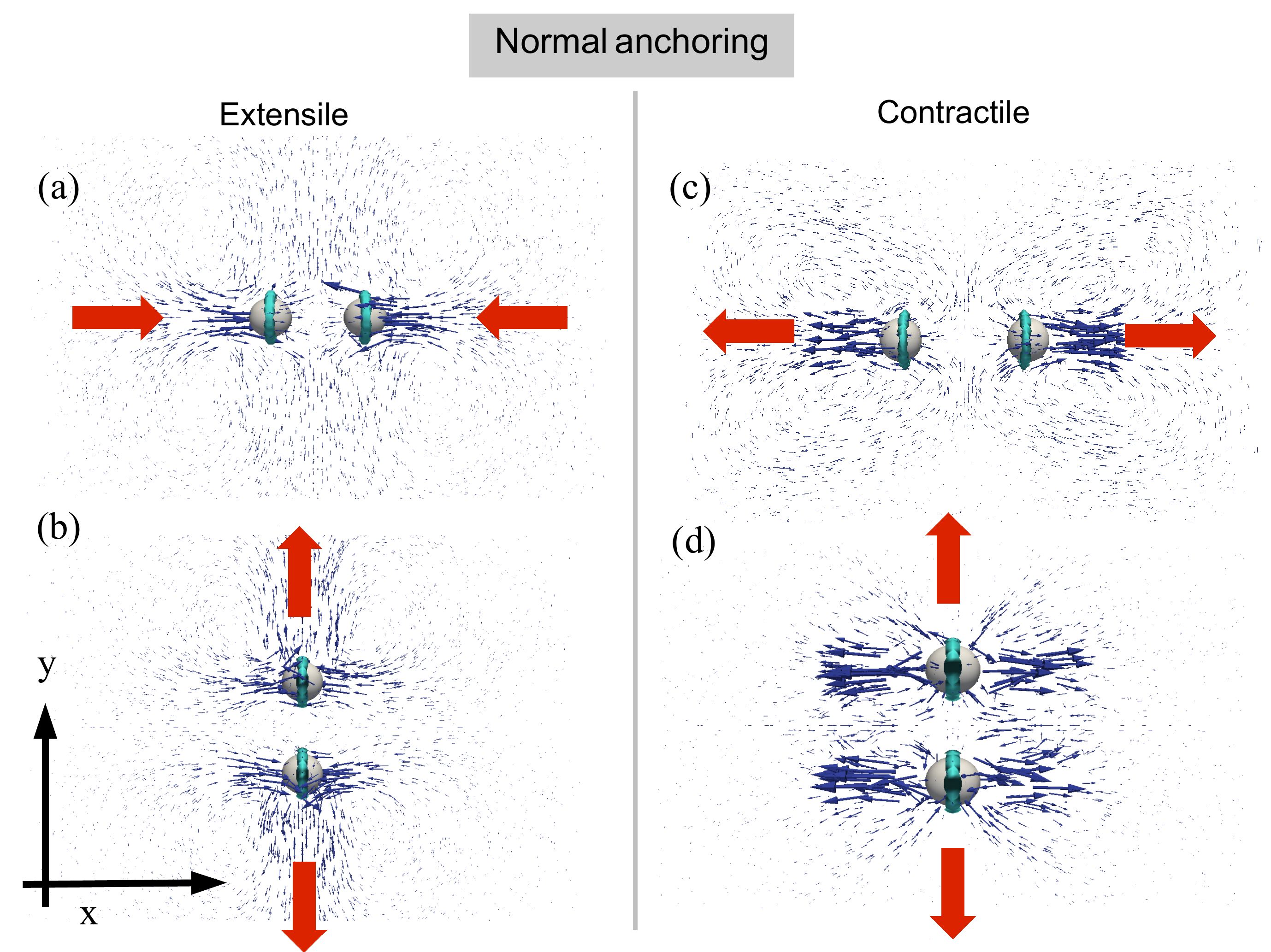}
\includegraphics[width=0.8\linewidth]{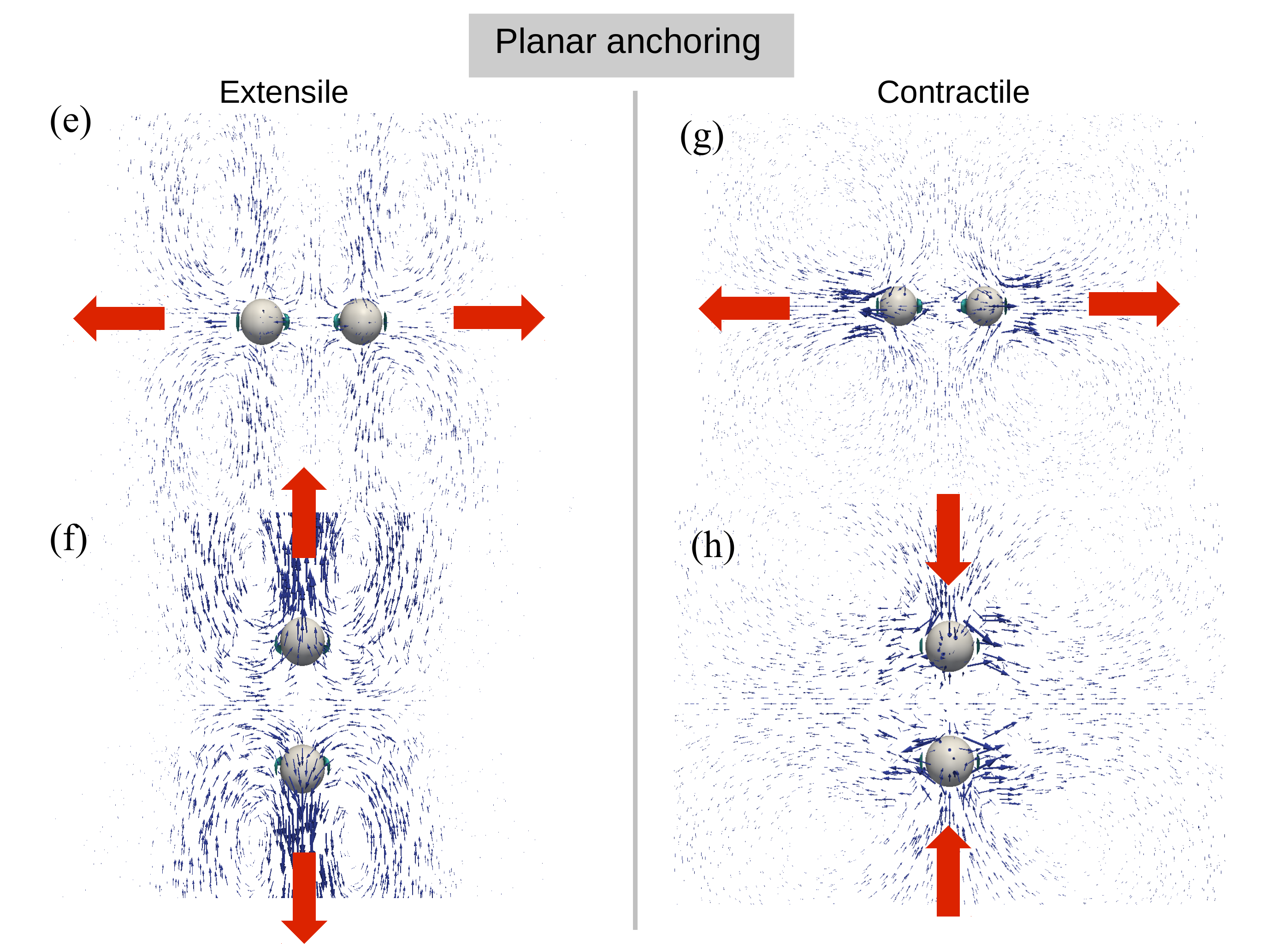}
\caption{Snapshots of two colloidal particles in extensile ($\xi=0.005$) and contractile ($\xi=-0.01$) nematic gels. The red arrows denote the direction of the particle movement. Equilibrium nematic direction is horizontal along the page.
For each case we show the velocity field close to the colloidal particles, and we indicate by red arrows whether we observed repulsion or attraction. 
The colloids attracted only in two of the eight cases considered: (1) in extensile fluid when the dimer is along the nematic director (Fig.~\ref{fig:2-SI}(a)) 
and (2) in contractile fluid when the dimer is perpendicular to director (Fig.~\ref{fig:2-SI}(h)). In all the other cases considered, the particles were 
driven away from each other by combination of elastic and advective forces.}
\label{fig:2-SI}
\end{figure}

\begin{figure*}[h]    
  \begin{center}
    \includegraphics[width=1.0\textwidth]{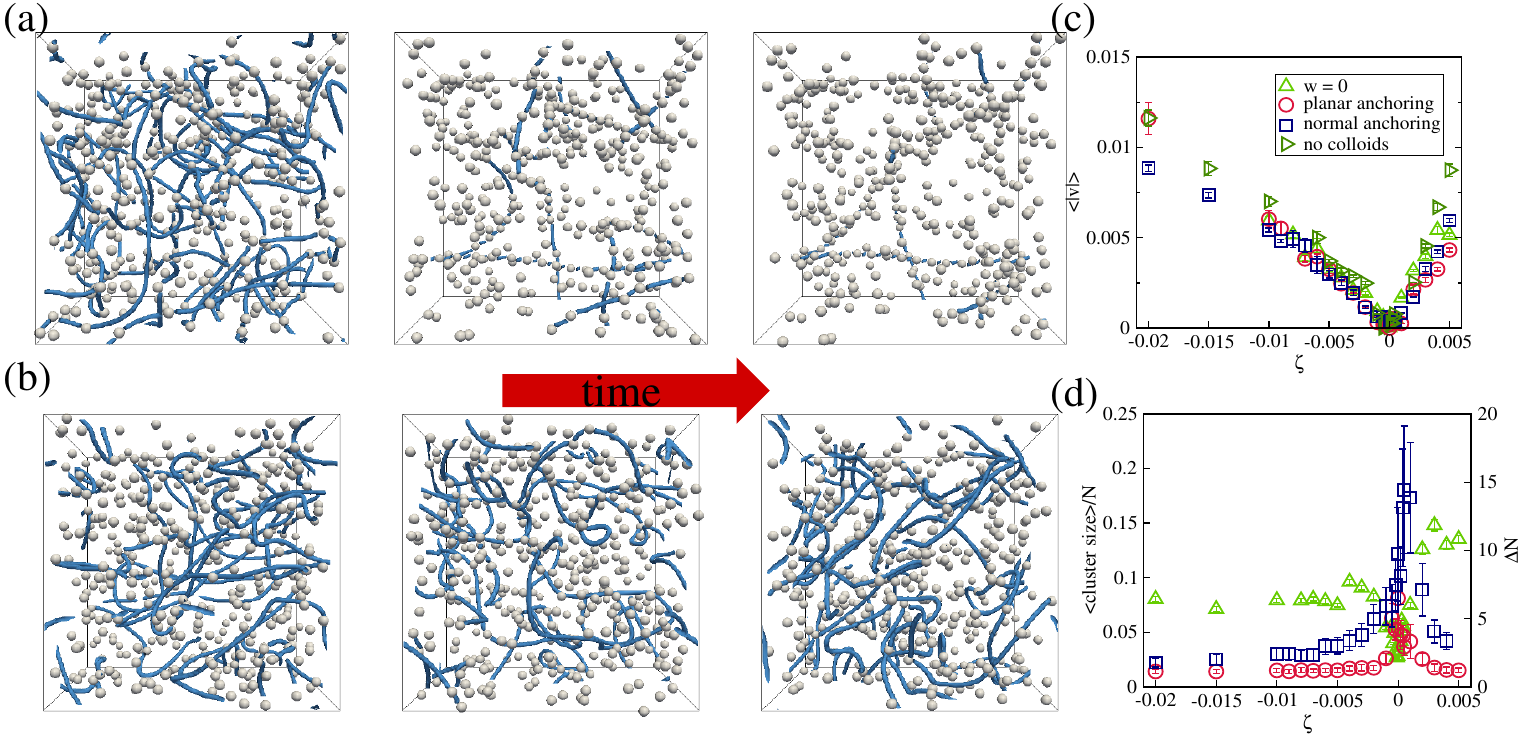}
  \end{center}
  \caption{
Snapshots of the $\varphi=15$\% colloid-active gel composite with $w=0$ following a quench from the isotropic to the nematic phase (set up in practice by starting from random director configuration): (a) passive fluid ($\zeta=0$) and (b) active nematic (extensile, $\zeta=0.005$). (The (gray) spheres are colloids and the (blue) ribbons are disclinations.) (c) Plot of the spatially average modulus of the fluid velocity, $\langle |v|\rangle$ as a function of $\zeta$. (d) Plot of the average colloid cluster size (for planar and normal anchoring) or of $\Delta N$ (for no anchoring), as a function of $\zeta$.}
\end{figure*}

We highlight two possible reasons why activity disrupts, rather than enhances, clustering, when $w>0$. First, it is well known that the spontaneous flow in active nematics has a turbulent and chaotic character~\cite{dogic,julia,cristina}: it can therefore be viewed as a nonequilibrium noise whose effective temperature increases with $\zeta$, and if large it can destabilise the thermodynamic interactions promoting clustering. (This is only a simplified view, as we have seen that such a noise cannot be considered a simple thermodynamic white noise, see Fig. 2(b).) Second, the interparticle attraction in the passive limit originates from the sharing of defects~\cite{tiffany,tanaka}: in a fully-developed spontaneously flowing regime, these tend to be more uniformly dispersed throughout the sample, rather than confined close to the particles, therefore gathering colloids now only brings about a limited thermodynamic benefit, screening the attraction which drives clustering. 
It should be noted that, with high enough $\zeta$, the behaviour should eventually become independent of $w$ (as active forces eventually dominate over anchoring), hence in this limit we would observe purely dynamic clustering due to advective forces, as in Fig. 2(b): therefore, the residual clustering in steady state observed with anchoring (Fig. 2(f)) will in general be due to a combination of activity-weakened thermodynamically driven clustering and activity-enhanced dynamic clustering.   

The previous results were obtained when starting from an orientationally ordered (monodomain) configuration of the active nematic. To address the role of different initial conditions, we disperse the particles within an isotropic host and quench the composite into the nematic phase -- this a setup that might be more closely connected to possible experiments (see below). 
 Fig. 5(a,b) shows snapshots from the dynamics at $w=0$. During the ordering process in the nematic following the quench, large disclination loops form in the bulk (Fig. 5(a,b)). These loops attract colloids, because covering the defects eliminates costly regions with high elastic energy density. When $\zeta=0$ the loops shrink, dragging colloids with them for part of the time; as a result there is still some mild residual clustering of the particles at the end of the process, see Fig. 5(a). (Because Brownian diffusion is not included in the simulations, particles remain where they have been left by the relaxing liquid crystal.) When $\zeta$ is large enough, spontaneous flow creates defects which persist in steady state (Fig. 5(b)), similary to what is observed with nematic initial conditions (see Fig. 1(d)), leading to dynamic clustering driven by nonequilibrium advective forces.

We further observe that, when performing a quench, the curves of $\langle |v| \rangle$ versus activity become largely independent of either the presence of colloids or the anchoring type, as is shown in Fig. 5(c). This is very different from what we found with the nematic initial condition (see Fig. 1). The reason is that following a quench the spontaneous flow mainly originates from the large elastic distortions and disclinations in the bulk, rather than from defects induced by the colloidal surfaces, as was the case with the nematic initial condition.
While the fluid component behaves quite differently in a quench, the colloidal behaviour is qualitatively similar to that observed with the nematic starting configuration (Fig. 2). Thus, for $w=0$ activity promotes dynamic clustering, whereas for $w>0$ it disrupts the colloidal aggregates which form in the passive system and re-distributes the particles more homogeneously (Fig. 5(d)). 

\section{Conclusions}

In conclusion, we have presented simulations of a {{quasi-2D}} concentrated suspension of passive colloidal particles in an active nematic gel.
The physics of the resulting ``inverse'' active composite material is at least as rich as that of its direct counterpart (active particles in a passive fluid), and results from a subtle interplay between activity and elasticity-induced interactions. 
Colloids whose surface imposes no orientational order on the active nematic dipoles, exhibit dynamic clustering, driven by the spontaneous flow induced by the activity in the bulk of the nematic host. {{The underlying physics is similar to that of fluctuation-dominated phase ordering~\cite{barma,barma2}, or path coalescence~\cite{mehlig,amos}, and is due to the fact that nearby particles are subjected to the same history of (chaotic) active flows, so that they end up in similar places.}}
On the other hand, when strong anchoring is imposed, activity provides a non-thermodynamic force which disrupts the aggregates which would otherwise form in the passive limit, and re-distributes the particles more uniformly.
Materials with tunable properties such as those studied here can be realised experimentally by dispersing colloids into extensile active gels formed by microtubule bundles and kinesin~\cite{dogic}, known to exhibit chaotic spontaneous flows powered by bend deformations in the microtubular orientation field. Another candidate composite material is a concentrated suspension of colloidal particles in actomyosin~\cite{actomyosin}. It would be interesting to study the emergent properties of such materials and compare them to those we find here. {{From a theoretical point of view, it would be of interest to study fully 3D colloid-active gel composites and assess whether the phenomenology found in our quasi-2D suspensions is retained there.}}

{\it Acknowledgements:} There are no conflicts of interest.

{\it Acknowledgements:} This work was funded by UK EPSRC grant EP/J007404/1, and in part by the European Research Council under the Horizon 2020 Programme, ERC grant agreement number 740269. JSL acknowledges EU intra-European fellowship 623637 DyCoCoS FP7-PEOPLE-2013-IEF, IdEx (Initiative d'Excellence) Bordeaux for funding. The authors would like to thank the Isaac Newton Institute for Mathematical Sciences for support and hospitality during the programme ``The mathematical design of new materials" when work on this paper was undertaken.

\appendix
\section{Appendix}
\subsection{Caption for Supplementary Movie 1}
This movie shows the full dynamics corresponding to the case of $w=0$ considered in Fig.~1(d), for large extensile activity. The arrows show the velocity field of the fluid. The movie shows that the chaotic character of spontaneous flow is similar to that observed in active nematics without colloids (see, e.g. ref.~\cite{julia}), 
It also demonstrates the dynamic clustering induced by the flow, which we describe in the text, and holds for no or weak anchoring (of the active nematic director at the colloidal surface).

\end{document}